\documentclass[pra,showpacs,preprintnumbers,amsmath,amssymb]{revtex4}
\usepackage{amsmath}
\usepackage{graphicx}% Include figure files
\usepackage{dcolumn}% Align table columns on decimal point
\usepackage{bm}% bold math
\usepackage{mathrsfs}
\usepackage{graphicx}
\usepackage{float}
\usepackage{color}
\usepackage{graphicx}
\usepackage{subfigure}
\begin{document}

%\preprint{APS/123-QED}
%\begin{CJK*}{GBK}{song}

\title{ Qubit-loss-free fusion of W states employing weak cross-Kerr nonlinearities}

\author {Meiyu Wang$^{1}$, Quanzhi Hao$^{1}$, Fengli Yan$^{1}$,}

\email{flyan@mail.hebtu.edu.cn}
\author {Ting Gao$^{2}$}
\email{gaoting@mail.hebtu.edu.cn}
\affiliation {$^{1}$College of Physics Science and Information Engineering, Hebei Normal University, Shijiazhuang 050024, China\\
$^{2}$ College of Mathematics and Information Science, Hebei Normal University, Shijiazhuang 050024, China}

\date{\today}

\begin{abstract}
{With the assistance of weak cross-Kerr nonlinearities, we introduce an optical scheme to fuse two small-size polarization entangled W states into a large-scale W state without qubit loss, i.e.,$\mathrm{W}_{n+m}$ state can be generated from an $n$-qubit W state and a $m$-qubit W state. To complete the fusion
task, two polarization entanglement processes and one spatial entanglement process are applied. The fulfillments of the above processes are contributed by a cross-Kerr nonlinear interaction between the signal photons and a coherent state via Kerr media.  We analyze the resource cost and the success probability of the scheme. There is no complete failure output in our fusion mechanism, and all the garbage states are recyclable. In addition, there is no need for any controlled quantum gate and any ancillary photon, so it is simple and feasible under the current experiment technology. }
\end{abstract}
%\end{CJK*}
\pacs{03.67.-a}
\maketitle

\section {Introduction}
 Entanglement \cite {Horodecki, YanGao1,YanGao2} plays an important role in quantum information processing (QIP), such as long-distance quantum communication and distributed quantum computation. Although most of the research in QIP are concerned with bipartite systems, multipartite entanglement has also attracted increasing interest. Compared with bipartite entangled states, multipartite entangled states have more complex and different entanglement structures. For a multipartite system, there exist many typical entangled states, including GHZ state, W state, cluster state, Dicke state, etc. Among the various multipartite entangled states, W states remain entanglement more robust. Exploiting it, quantum mechanics against local hidden variable theory can be tested \cite {Ou}. Moreover, W states are a necessary kind of physics resources in quantum teleportation and superdense coding \cite {Shi, Agrawal, Li1, Wang}, quantum deterministic secure comminication and key distribution \cite{Cao, Dong1}, optimal universal quantum cloning machine \cite{Murao}, as well as the leader election problem in anonymous quantum network \cite{Hondt}. Hence, to design simple and efficient scheme for preparing large-scale multipartite entangled W states is under intense research.

In recent years, expansion and fusion operations have been proposed as an efficient way to prepare large-scale multipartite entangled states. One can get a larger entangled state from two or more multipartite entangled states by sending only one qubit of each seed entangled state to the fusion operation.  For instance, efficient preparation and expansion of GHZ and cluster states are well known \cite {Browne, Zeilinger}. The creation of W states via the fusion process have attracted considerable attention \cite {Zou, Yamamoto, Xiang, Tashima1, Tashima2, Ozdemir, Bugu, Yesilyurt, Ozaydin, Han, Li} too. In 2011, \"{O}zdemir \emph{et al.} first proposed an optical fusion scheme for W states, with which a $\mathrm{W}_{n+m-2}$ state can be generated from $\mathrm{W}_{n}$ and $\mathrm{W}_{m}$ ($n,m \geq 3$) states \cite{Ozdemir}. In 2013, Bugu \emph{et al.} \cite{Bugu} made a great improvement on the basis of the scheme in Ref. \cite{Ozdemir}, which can achieve a $\mathrm{W}_{n+m-1}$ from $\mathrm{W}_{n}$ and $\mathrm{W}_{m}$ ($n,m \geq 2$) with the help of a single Fredkin gate and an ancillary photon. Subsequently, other two schemes in Refs. \cite {Yesilyurt, Ozaydin} were proposed for fusing W states. The similarity of the schemes \cite { Bugu, Yesilyurt, Ozaydin} is to introduce controlled quantum gates and
ancillary qubits to enhance the efficiency of the fusion mechanism. However, it is a great challenge to realize these quantum logic gates with current experimental technology, which will increase the realization complexity of the fusion process. In 2015, Han \emph{et al.} proposed a new scheme to fuse an \emph{n}-qubit W state and a \emph{m}-qubit W state with the weak cross-Kerr nonlinearities \cite {Han}, and an $(n + m - 1$)-qubit W state can be generated without any ancillary photon. It is worth pointing out that one or two of the qubits entering the fusion mechanism must be measured so as to complete the whole fusion process, i.e., there is qubit loss in most of the previous works. As a result, the number of the output entangled qubits is smaller than the sum of numbers of the input entangled qubits, which will inevitably decrease the fusion efficiency and increase the number of fusion steps. A new scheme was proposed to solve this problem in Ref. \cite{Li}. The scheme \cite{Li} designed a fusion mechanism to fuse two small-size W states into a large-scale W state without qubit loss, and it is called qubit-loss-free (QLF) fusion mechanism, which is based on a two-outcome positive-operator valued measurement on two qubits extracting from two small-size W states.

In this paper, we propose an alternative scheme that can obtain an $(n + m$)-qubit W state by fusing an $n$-qubit W state and $m$-qubit W state $(n,m \geq 2)$ with the QLF fusion mechanism following some ideas in Ref. \cite {Li}. Different from the Ref. \cite {Li}, we fulfill the fusion process in optical system based on weak cross-Kerr nonlinearities. The whole fusion scheme can be separated into three processes, which are two polarization entanglement processes of two photons coming from each seed entangled state and a spatial entanglement process. The paper is organized as follows: in Sec. II, we concretely construct the setups
to fuse the W state. The resource cost and  the experimental feasibility of the scheme are analyzed in Sec. III. Finally, our main work are summarized in Sec. IV.

\section { Qubit-loss-free fusion mechanism for W state with cross-Kerr nonlinearities }

 The cross-Kerr nonlinearity can be described
with the Hamiltonian $\hat{H}_{k}=-\hbar\kappa\hat{n}_{s}\hat{n}_{p}$,
where $\hat{n}_{s}$ $(\hat{n}_{p}) $ is the photon-number operator of
the signal (probe) mode, and $\kappa$ is the strength
of the nonlinearity. If the signal field contains $n$ photons and the probe field is
 in an initial coherent state with  amplitude $\alpha$, the
cross-Kerr nonlinearity interaction causes the combined signal-probe
system to evolve as follows:
\begin{equation}
\mathrm{e}^{-\mathrm{i}\hat{H}_{k}t/\hbar}|n\rangle_{s}|\alpha\rangle_{p}=\mathrm{e}^{\mathrm{i}\kappa t\hat{n}_{s}\hat{n}_{p}}|n\rangle_{s}|\alpha\rangle_{p}=|n\rangle_{s}|\alpha\mathrm{e}^{\mathrm{i}n\theta}\rangle_{p},
\end{equation}
where $\theta=\kappa t$ with $t$ being the
interaction time. It is easy to observe that signal-photon state is unaffected by the
interaction but the coherent state picks up a phase shift $n\theta$ directly
proportional to the number of photons $n$ in the signal mode.
One can exactly obtain the information of photons in the signal state
but not destroy them through a general homodyne-heterodyne measurement of the phase of the coherent state. This technique was first
used to realize a parity  gate \cite {Barrett} then a CNOT gate
\cite {Nemoto}, where the requirement for this technique is $ \alpha\theta^{2} > 9$ with $\alpha$ denoting
the amplitude of the coherent state.  As for the cross-Kerr nonlinearity, the nonlinearity magnitude $\theta\sim10^{-2}$ are potentially available with the help of electromagnetically induced transparency \cite{Munro1}. In particular, the error probability is $P_{\texttt{error}}=3.4\times10^{-6}$ on the condition $\alpha=90000, \theta=0.01$. This shows that it is still possible to operate in the regime of weak cross-Kerr nonlinearity, and the amplitude of the probe coherent state beam is physical reasonable with current experimental technology.
%In what follows, we will explain the detailed QLF fusion procedures for generating the large-scale maximally entangled W states by fusing two small-size entangled W states.

The fusion scenario we proposed is as follows. Suppose Alice and Bob possess $n$- and $m$-qubit polarization entangled W states, $|\mathrm{W}_{n}\rangle$ and
$|\mathrm{W}_{m}\rangle$ , respectively. They wish to fuse their states to obtain a large-scale W state with the help of the weak cross-Kerr nonlinearities by sending only one photon from their states to the scheme shown in Fig.1. We denote the polarization entangled states of Alice and Bob as
\begin{eqnarray}\label{2}
|\mathrm{W}_{n}\rangle_{\mathrm{A}}&=&\frac{1}{\sqrt{n}}[|(n-1)_{H}\rangle_{\mathrm{a}}|1_{V}\rangle_{1}+\sqrt{n-1}|\mathrm{W}_{n-1}\rangle_{\mathrm{a}}|1_{H}\rangle_{1}],\\
|\mathrm{W}_{m}\rangle_{\mathrm{B}}&=&\frac{1}{\sqrt{m}}[|(m-1)_{H}\rangle_{\mathrm{b}}|1_{V}\rangle_{2}+\sqrt{m-1}|\mathrm{W}_{m-1}\rangle_{\mathrm{b}}|1_{H}\rangle_{2}],
\end{eqnarray}
where the photons in modes 1 and 2 can be accessed from each W state and those in modes a and b are kept intact at their site. The direct product of the
two input W states can be written as
\begin{alignat}{2}
|\mathrm{W}_{n}\rangle_{\mathrm{A}}\otimes|\mathrm{W}_{m}\rangle_{\mathrm{B}}=&\frac{1}{\sqrt{nm}}|(n-1)_{H}\rangle_{\mathrm{a}}|(m-1)_{H}\rangle_{\mathrm{b}}|1_{V}\rangle_{1}|1_{V}\rangle_{2}
+\frac{\sqrt{n-1}}{\sqrt{nm}}|\mathrm{W}_{n-1}\rangle_{\mathrm{a}}|(m-1)_{H}\rangle_{\mathrm{b}}|1_{H}\rangle_{1}|1_{V}\rangle_{2}\nonumber\\
&+\frac{\sqrt{m-1}}{\sqrt{nm}}|(n-1)_{H}\rangle_{\mathrm{a}}|\mathrm{W}_{m-1}\rangle_{\mathrm{b}}|1_{V}\rangle_{1}|1_{H}\rangle_{2}
 +\frac{\sqrt{(n-1)(m-1)}}{\sqrt{nm}}|\mathrm{W}_{n-1}\rangle_{\mathrm{a}}|\mathrm{W}_{m-1}\rangle_{\mathrm{b}}|1_{H}\rangle_{1}|1_{H}\rangle_{2}\nonumber\\
=&|a\rangle|1_{V}\rangle_{1}|1_{V}\rangle_{2}+|b\rangle|1_{H}\rangle_{1}|1_{V}\rangle_{2}+|c\rangle|1_{V}\rangle_{1}|1_{H}\rangle_{2}+|d\rangle|1_{H}\rangle_{1}|1_{H}\rangle_{2},
\end{alignat}
for convenience, where we have substituted $|a\rangle$ for $\frac{1}{\sqrt{nm}}|(n-1)_{H}\rangle_{\mathrm{a}}|(m-1)_{H}\rangle_{\mathrm{b}}$, $|b\rangle$ for $\frac{\sqrt{n-1}}{\sqrt{nm}}|\mathrm{W}_{n-1}\rangle_{\mathrm{a}}|(m-1)_{H}\rangle_{\mathrm{b}}$,  $|c\rangle$ for $\frac{\sqrt{m-1}}{\sqrt{nm}}|(n-1)_{H}\rangle_{\mathrm{a}}|\mathrm{W}_{m-1}\rangle_{\mathrm{b}}$, and $|d\rangle$ for $\frac{\sqrt{(n-1)(m-1)}}{\sqrt{nm}}|\mathrm{W}_{n-1}\rangle_{\mathrm{a}}|\mathrm{W}_{m-1}\rangle_{\mathrm{b}}$.

According to the scheme shown in Fig.1, the fusion mechanism mainly consists of three steps. In the first step, the photons in modes 1 and 2 pass through the $\mathrm{PBS_{1}}$ and $\mathrm{PBS_{2}}$, and then interact with the coherent probe beam via the cross-Kerr nonlinear medium. The action of the $\mathrm{PBS_{s}}$, cross-Kerr nonlinearity and a further linear phase shift will evolve the joint state of the combined system $|\mathrm{W}_{n}\rangle_{\mathrm{A}}|\mathrm{W}_{m}\rangle_{\mathrm{B}}|\alpha\rangle$ to
\begin{eqnarray}
|a\rangle|1_{V}\rangle_{1}|1_{V}\rangle_{2}|\alpha\mathrm{e}^{\frac{1}{2}\mathrm{i}\theta}\rangle+|b\rangle|1_{H}\rangle_{1}|1_{V}\rangle_{2}|\alpha\mathrm{e}^{-\frac{1}{2}\mathrm{i}\theta}\rangle
+|c\rangle|1_{V}\rangle_{1}|1_{H}\rangle_{2}|\alpha\mathrm{e}^{-\frac{1}{2}\mathrm{i}\theta}\rangle+|d\rangle|1_{H}\rangle_{1}|1_{H}\rangle_{2}|\alpha\mathrm{e}^{-\frac{3}{2}\mathrm{i}\theta}\rangle.
\end{eqnarray}

\begin{figure}
\begin{center}
\scalebox{0.8}{\includegraphics* {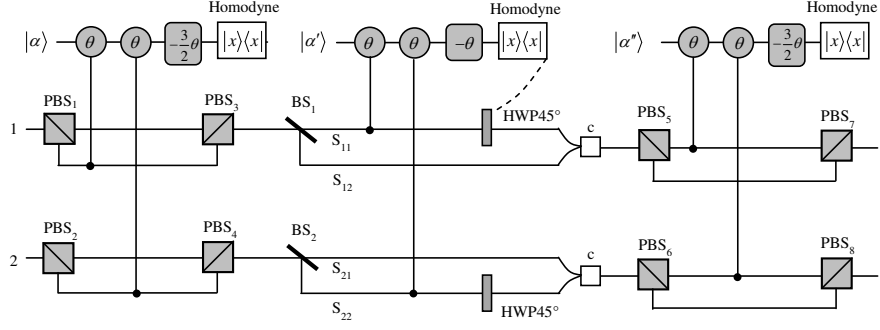}}
\caption{An illustration plot for QLF fusion mechanism. The polarization beam splitters (PBSs) distinguish horizontal polarization states $|H\rangle$ and
vertical polarization states $|V\rangle$ and allow them to go through different lines. The beam splitters (BSs) have equal probabilities (50:50) of transmission and reflection. Three pairs of  phase shift $\theta$ are accumulated on the coherent states $|\alpha\rangle$, $|\alpha'\rangle$,$|\alpha''\rangle$ respectively after undergoing cross-Kerr nonlinear interaction. Before entering into the measurement setup $|x\rangle\langle x|$, the phase modulation $-\frac{3\theta}{2}$ (or $-\theta$ ) is employed to change the phase of the corresponding coherent state. Half-wave plate, HWP45, realize single photon $\sigma_{x}$ operation. C is the path coupler to combine the two paths of each photon, which acts as a quantum eraser to erase the path information without destroying the polarization information.}
\end{center}
\end{figure}

Two scenarios of phase shifts $\pm\frac{1}{2}\mathrm{i}\theta$ and phase shift $-\frac{3}{2}\mathrm{i}\theta$ that occurred on the coherent state $|\alpha\rangle$ need to be distinguished, which can be realized by an X-quadrature homodyne measurement \cite{Barrett} on the coherent state. If $-\frac{3}{2}\mathrm{i}\theta$ phase shift is witnessed, the state $|d\rangle|1_{H}\rangle_{1}|1_{H}\rangle_{2}$ can be achieved. In this case, the remaining parts at the sides of Alice and Bob are still W states with a smaller number of qubits, which can be used to fuse again. If phase shifts $\pm\frac{1}{2}\mathrm{i}\theta$ are obtained, a phase shift $2\phi(x,\frac{\theta}{2})$ operation should be performed on horizontal polarization components to erase the phase difference. Here $2\phi(x,\frac{\theta}{2})=2\alpha\sin \frac{\theta}{2} (x-2\alpha \cos \frac{\theta}{2}) \mathrm{mod}$  $2\pi$ is a function of the phase shift and the eigenvalue $x$ of the X operator. Then the following state is obtained with the success probability $\frac{n+m-1}{nm}$
\begin{eqnarray}
|\Phi\rangle\sim|a\rangle|1_{V}\rangle_{1}|1_{V}\rangle_{2}+|b\rangle|1_{H}\rangle_{1}|1_{V}\rangle_{2}+|c\rangle|1_{V}\rangle_{1}|1_{H}\rangle_{2}.
\end{eqnarray}

In the second step, the state $|\Phi\rangle$ is used to continue the fusion process. In the spatial entanglement gate, the two photons pass through beam
splitters BS$_{1}$ and BS$_{2}$, which have the following function between two input  modes (a,b) and two output modes (c,d): $a^{\dagger}\rightarrow(c^{\dagger}+d^{\dagger})/\sqrt{2}, b^{\dagger}\rightarrow(c^{\dagger}-d^{\dagger})/\sqrt{2}$,
  the photons (1,2) enter into the  paths $(\mathrm{S}_{11}, \mathrm{S}_{12}$) and the paths ($\mathrm{S}_{21}, \mathrm{S}_{22}$) respectively. Accompanying with the coherent state, the photons (1,2) enter into Kerr media. Then, the state of photons (1,2) with the coherent state $|\alpha'\rangle$ evolves as
\begin{eqnarray}
|\Phi\rangle|\alpha'\rangle&\rightarrow&\frac{1}{2}(|a\rangle|1_{V}\rangle_{1}|1_{V}\rangle_{2}+|b\rangle|1_{H}\rangle_{1}|1_{V}\rangle_{2}+|c\rangle|1_{V}\rangle_{1}|1_{H}\rangle_{2})(|\mathrm{S}_{11}\rangle|\mathrm{S}_{21}\rangle+|\mathrm{S}_{12}\rangle|\mathrm{S}_{22}\rangle)|\alpha'\rangle\nonumber\\
&&+\frac{1}{2}(|a\rangle|1_{V}\rangle_{1}|1_{V}\rangle_{2}+|b\rangle|1_{H}\rangle_{1}|1_{V}\rangle_{2}+|c\rangle|1_{V}\rangle_{1}|1_{H}\rangle_{2})(|\mathrm{S}_{11}\rangle|\mathrm{S}_{22}\rangle|\alpha'\mathrm{e}^{\mathrm{i}\theta}\rangle+|\mathrm{S}_{12}\rangle|\mathrm{S}_{21}\rangle|\alpha'\mathrm{e}^{-\mathrm{i}\theta}\rangle).
\end{eqnarray}
Performing an X homodyne measurement on the coherent state with $\alpha'$ real, there are two measurement outcomes
corresponding to scenarios of phase shift ($0,\pm\theta$). Explicitly, if zero phase shift occurs, the state will be projected into
\begin{eqnarray}
|\Psi\rangle\sim\frac{1}{2}(|a\rangle|1_{V}\rangle_{1}|1_{V}\rangle_{2}+|b\rangle|1_{H}\rangle_{1}|1_{V}\rangle_{2}+|c\rangle|1_{V}\rangle_{1}|1_{H}\rangle_{2})(|\mathrm{S}_{11}\rangle|\mathrm{S}_{21}\rangle+|\mathrm{S}_{12}\rangle|\mathrm{S}_{22}\rangle).
\end{eqnarray}
 Otherwise, nonzero phase shift is presented, the photons are in the following state
 \begin{eqnarray}
|\Psi'\rangle\sim\frac{1}{2}(|a\rangle|1_{V}\rangle_{1}|1_{V}\rangle_{2}+|b\rangle|1_{H}\rangle_{1}|1_{V}\rangle_{2}+|c\rangle|1_{V}\rangle_{1}|1_{H}\rangle_{2})(|\mathrm{S}_{11}\rangle|\mathrm{S}_{22}\rangle+|\mathrm{S}_{12}\rangle|\mathrm{S}_{21}\rangle).
\end{eqnarray}
It is worth noting that the state denoted as Eq. (9) is the same as Eq. (8) when a swap gate is inserted into the paths $S_{21}$ and $S_{22}$.  A swap gate is an important two-qubit logic gate. In terms of the basis of $\{|00\rangle, |01\rangle, |10\rangle, |11\rangle\}$, the swap gate can be represented as the following matrix:
\begin{gather}
\begin{bmatrix}1 & 0 & 0 & 0\\0 & 0 & 1 & 0\\0 & 1 & 0 & 0\\0 & 0 & 0 & 1\end{bmatrix}.
\end{gather}
In practice, the swap gate transformation can be yielded by the Hong-Ou-Mandel interference \cite{Hong} in the Mach-Zehnder interferometer \cite{Lin1, Milburn}, illustrated in Fig.2. Two beam splitters constitute a Mach-Zehnder interferometer. Additionally, the phase shifter PS $\pi$ denotes the phase shift $\pi$ executed on the photon passing through the line it is inserted.

For simplifying description in the later process, we only consider the case of zero phase shift. If zero phase shift is witnessed by the X homodyne measurement, half wave plates, HWP45¡ãs, are inserted into the paths $S_{11}$ and $S_{22}$ at first to perform $\sigma_{x}$ operation. Then, each path coupler C can combine the two paths of corresponding photon, which acts as a quantum eraser to erase the path information of the photon  without destroying the polarization information. Therefore, the state in Eq.(8) will be
\begin{eqnarray}
|\Psi\rangle\propto\frac{1}{2}[|a\rangle(|1_{H}\rangle_{1}|1_{V}\rangle_{2}+|1_{V}\rangle_{1}|1_{H}\rangle_{2})
+|b\rangle(|1_{V}\rangle_{1}|1_{V}\rangle_{2}+|1_{H}\rangle_{1}|1_{H}\rangle_{2})
+|c\rangle(|1_{H}\rangle_{1}|1_{H}\rangle_{2}+|1_{V}\rangle_{1}|1_{V}\rangle_{2})].
\end{eqnarray}

In the third step, the photons 1 and 2 enter into the second polarization entanglement gate. As the consequence of the nonlinear interaction between photons and the
coherent state, the state of the whole system can be expressed as
\begin{eqnarray}
|\Psi\rangle|\alpha''\rangle&\rightarrow&\frac{1}{2}[|a\rangle(|1_{H}\rangle_{1}|1_{V}\rangle_{2}+|1_{V}\rangle_{1}|1_{H}\rangle_{2})|\alpha''\mathrm{e}^{\mathrm{-i}\frac{\theta}{2}}\rangle
+|b\rangle|1_{H}\rangle_{1}|1_{H}\rangle_{2}|\alpha''\mathrm{e}^{\mathrm{i}\frac{\theta}{2}}\rangle
+|c\rangle|1_{H}\rangle_{1}|1_{H}\rangle_{2})|\alpha''\mathrm{e}^{\mathrm{i}\frac{\theta}{2}}\rangle\nonumber\\
&&+|b\rangle|1_{V}\rangle_{1}|1_{V}\rangle_{2}|\alpha''\mathrm{e}^{\mathrm{-i}\frac{3\theta}{2}}\rangle
+|c\rangle|1_{V}\rangle_{1}|1_{V}\rangle_{2})|\alpha''\mathrm{e}^{\mathrm{-i}\frac{3\theta}{2}}\rangle].
\end{eqnarray}

After performing an X homodyne measurement on the coherent state, two scenarios of phase shifts $\pm\frac{1}{2}\mathrm{i}\theta$ and phase shift $-\mathrm{i}\frac{3\theta}{2}$  are occurred. If phase shifts $\pm\frac{1}{2}\mathrm{i}\theta$ are obtained, photons are in the following state

\begin{eqnarray}
|\mathrm{W}'\rangle&\sim&\frac{1}{2}[|a\rangle(|1_{H}\rangle_{1}|1_{V}\rangle_{2}+|1_{V}\rangle_{1}|1_{H}\rangle_{2})
+|b\rangle|1_{H}\rangle_{1}|1_{H}\rangle_{2}+|c\rangle|1_{H}\rangle_{1}|1_{H}\rangle_{2}]\nonumber\\
&=&\frac{1}{\sqrt{2}}[\frac{1}{\sqrt{nm}}|(n-1)_{H}\rangle_{\mathrm{a}}|(m-1)_{H}\rangle_{\mathrm{b}}|\mathrm{W}_{2}\rangle_{12}
+\frac{\sqrt{n-1}}{\sqrt{2nm}}|\mathrm{W}_{n-1}\rangle_{\mathrm{a}}|(m-1)_{H}\rangle_{\mathrm{b}}|1_{H}1_{H}\rangle_{12}\nonumber\\
&&+\frac{\sqrt{m-1}}{\sqrt{2nm}}|(n-1)_{H}\rangle_{\mathrm{a}}|\mathrm{W}_{m-1}\rangle_{\mathrm{b}}|1_{H}1_{H}\rangle_{12}]\nonumber\\
&=&\frac{\sqrt{n+m}}{2\sqrt{nm}}|\mathrm{W}_{n+m}\rangle.
\end{eqnarray}

If $-\mathrm{i}\frac{3\theta}{2}$  phase shift is witnessed, Von Neumann projection measurements on photons 1 and 2 can be made, the $(n + m -2)$
qubits will be left in the state $|\mathrm{W}_{n+m-2}\rangle$, which is a large-scale W state, too. Moreover, if the $|\mathrm{W}_{2}\rangle$ state is available, a
$|\mathrm{W}_{n+m-2}\rangle$state and a $|\mathrm{W}_{2}\rangle$ state can be fused into an $(n + m)$-qubit maximally entangled W state by our QLF fusion scheme. Furthermore, after the X homodyne measurement in the first step, although the state $|d\rangle|1_{H}\rangle_{1}|1_{H}\rangle_{2}$ is achieved, the remaining qubits of each of the $W$ states still keep their entanglement structure intact so that a new round of fusion can be performed on them.  Availability of these two recyclable cases
may increase the efficiency of the process and reduce the cost of preparing the desired state.

\begin{figure}
\begin{center}
\scalebox{0.8}{\includegraphics* {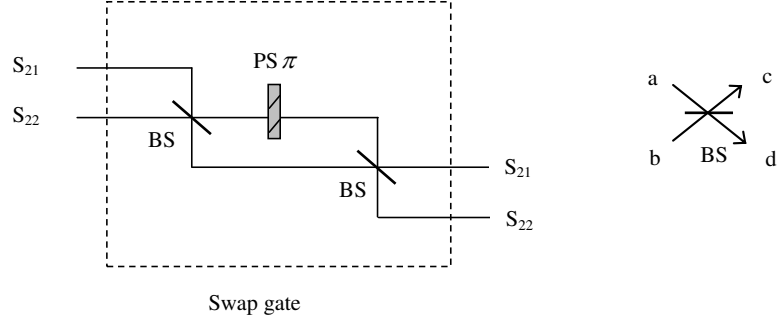}}
\caption{Illustration plot for depicting the swap gate. The symbol PS $\pi$ denotes the phase shift $\pi$ executed on the photon passing through the line it is inserted. A beam splitter has the following function between two input modes (a, b) and two output modes (c, d): $a^{\dagger}\rightarrow \frac{1}{\sqrt{2}}(c^{\dagger}+d^{\dagger})$, $b^{\dagger}\rightarrow \frac{1}{\sqrt{2}}(c^{\dagger}-d^{\dagger})$.}
\end{center}
\end{figure}

\begin{figure}
\begin{center}
\subfigure{ \label{fig:subfig:a}}
\scalebox{0.3}{\includegraphics* {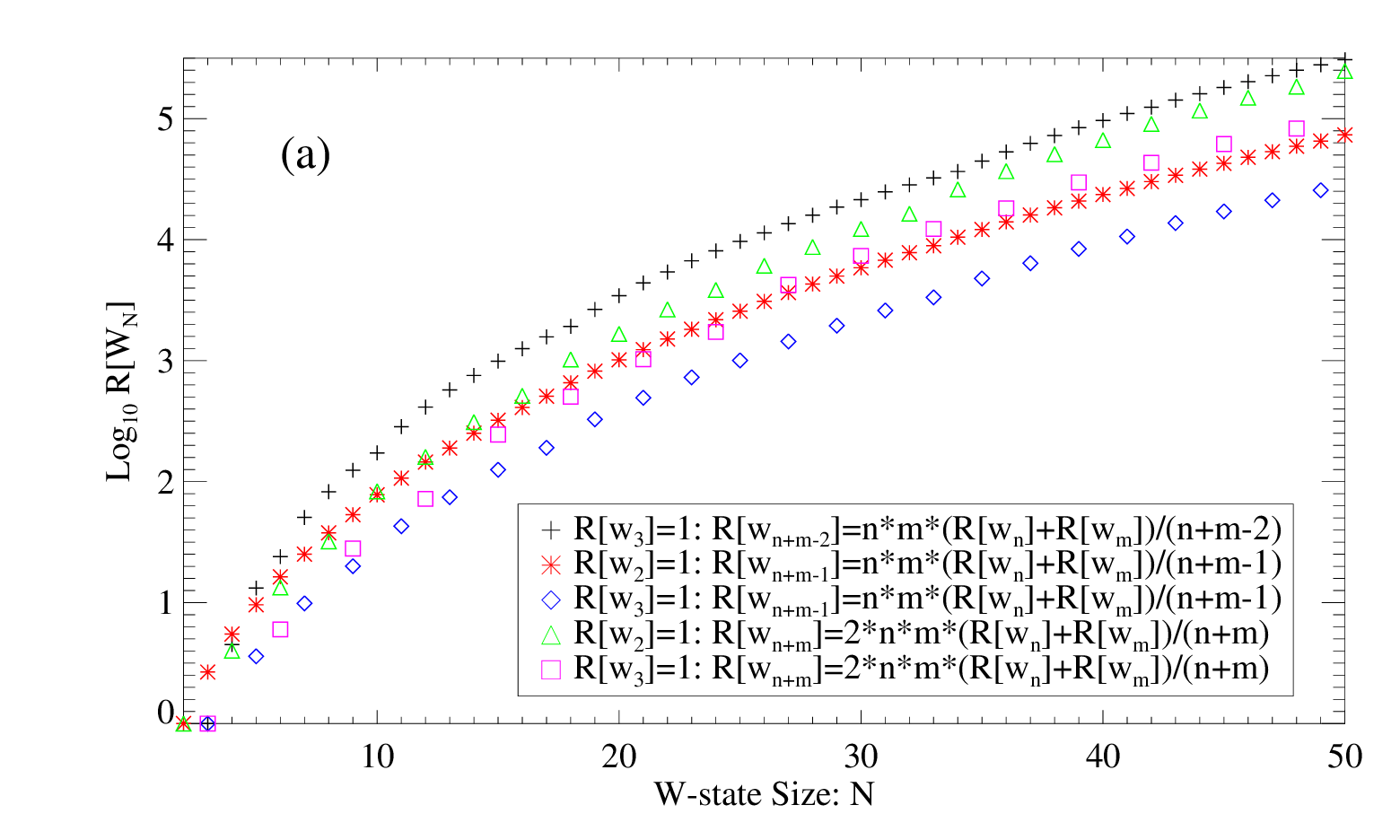}}
\subfigure{\label{fig:subfig:b}}
\scalebox{0.3}{\includegraphics* {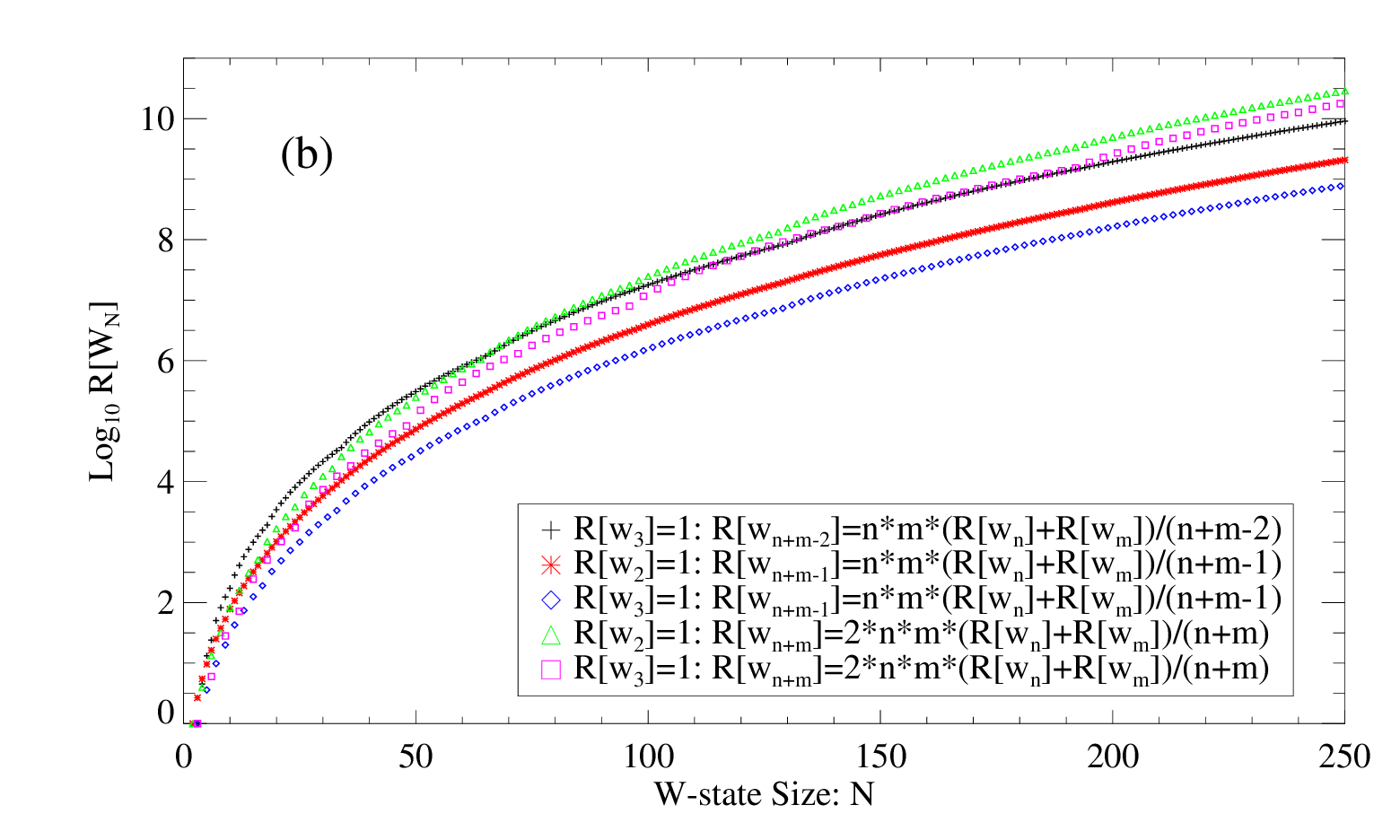}}
\caption{The optimal cost of comparison: the fusion scheme in [20](black $+$), the fusion mechanism in [21] in  which the initial state is $\mathrm{W}_{2}$ (red $*$) and the initial state is $\mathrm{W}_{3}$ (blue rhombus) respectively, our fusion scheme in the case of the initial state $\mathrm{W}_{2}$ (green triangle) and $\mathrm{W}_{3}$ (purple square) respectively. In Fig. 3a, the size of the fusion W  state is under 50, while in Fig. 3b, the size of the fusion $W$ state is under 250.}

\end{center}
\end{figure}

\section{ analysis and discussion}
In this section, we will estimate the performance of the QLF fusion scheme by analyzing the resource cost and feasibility of preparing large-size W states.
 Similar to Ref. \cite{Ozdemir}, we use the notation $R[\mathrm{W}_{n+m}]$ to denote the resource cost of creating state $\mathrm{W}_{n+m}$, which is defined as
\begin{equation}
R[\mathrm{W}_{n+m}]=\frac{R[\mathrm{W}_{n}]+R[\mathrm{W}_{m}]}{P_{s}(\mathrm{W}_{n},\mathrm{W}_{m})},
\end{equation}
where $P_{s}(\mathrm{W}_{n},\mathrm{W}_{m})$ is the success probability for fusing a $\mathrm{W}_{n}$ and a $\mathrm{W}_{m}$ into a $\mathrm{W}_{n+m}$. From the whole fusion process, we can see that the fused large W state is first from the items of the initial state with the photons 1 and 2 in the states $|1_{V}\rangle_{1}|1_{V}\rangle_{2}$, $|1_{V}\rangle_{1}|1_{H}\rangle_{2}$, $|1_{H}\rangle_{1}|1_{V}\rangle_{2}$ with success probability $P_{s}(1)= \frac{n+m-1}{nm}$. In the third step, the $\mathrm{W}_{n+m}$ state can be obtained with success probability $P_{s}(2)= \frac{n+m}{2(n+m-1)}$. Therefore, the total success probability
$P_{s}$ is written as
\begin{equation}
P_{s}=P_{s}(1)\times P_{s}(2)= \frac{n+m-1}{nm}\times\frac{n+m}{2(n+m-1)}=\frac{n+m}{2nm},
\end{equation}
which means we have the same entangled-resource cost with Ref. \cite {Li}.

The numerical results of the optimal costs of creating state $\mathrm{W}_{n+m}$ for three fusion schemes (including our QLF fusion scheme) are shown in Fig. 3. Since the Ref. \cite {Bugu} and our QLF fusion scheme can work for bipartite Bell state $\mathrm{W}_{2}$ , the numerical results of the resource costs with the basic resource $R[\mathrm{W}_{2}]=1$ and $R[\mathrm{W}_{2}]=1$ are also shown in Fig. 3 respectively.  The comparison will be done with the fusion schemes reported in Refs. \cite {Ozdemir} and \cite {Bugu}. Reference \cite {Ozdemir} clearly states that the optimal cost of any state can be numerically calculated using the recursive formula
\begin{equation}
R[\mathrm{W}_{n}]_{opt}=\min\frac{R[\mathrm{W}_{k}]_{opt}+R[\mathrm{W}_{n-k}]_{opt}}{P_{s}(\mathrm{W}_{k},\mathrm{W}_{n-k})},
\end{equation}
and it indicates the closer the sizes of the two resource states are, the lower the cost is. Our resource cost analysis is also based on this principle. From this figure, we can see that for creating W states with the same size, the resource cost using $\mathrm{W}_{3}$ as initial resource is lower than the one using $\mathrm{W}_{2}$ both in Ref. \cite {Bugu} and our QLF fusion scheme. On the other hand, it can be seen that when the size of the fused W state is not so large, for instance $N=50$, our QLF fusion scheme is more efficient than the scheme of Ref. \cite {Ozdemir}. Although Ref. \cite {Bugu} is always more efficient than our fusion scheme, as mentioned before, a Fredkin gate and an ancillary qubit have been introduced to enhance the efficiency of the fusion mechanism in Ref. \cite {Bugu}, but the implementation of a Fredkin gate is not an easy task, which is difficult to be realized in practical experiment.

 Now we give a brief discussion about the experimental feasibility of protocol with the current experimental technology.
On the one hand, the cross-Kerr nonlinearity is an important part of the present scheme.  All the nonlinearities required in our scheme have the same nonlinear strength, which can reduce the difficulty in experiment. In the fusion process, three X homodyne measurements may introduce errors to the prepared large-size W states. From the beginning part of section II, we can see the error probability of the X homodyne measurement lies on the value of $\alpha\theta^{2}$ and decreases as $\alpha\theta^{2}$ increases.  When $\alpha\theta^{2}>9$, the error probability is of an order less than $10^{-5}$, which implies that the discrimination is desirable in the optical regime. So the error probability introduced by X homodyne measurements is ignored in calculating the total success probability. On the other hand, we should take into account of the effect of decoherence in the transmission
of the coherence probe beam. In the real situation, photon loss or amplitude damping is the main
source of this kind decoherence. When decoherence occurs, the pure $\mathrm{W}_{n}$ and $\mathrm{W}_{m}$ states will
evolve to mixed states and its fidelity will decrease. Fortunately,
the decoherence can be made arbitrarily small simply by an arbitrary strong coherent
state associated with a displacement operation on the coherent state and the QND
photon-number-resolving detection \cite{Jeong,Barrett1}. An earlier analysis was given by Munro et al. \cite{Munro}. We think the related analysis of against small loss of photons can also apply to our scheme in a way.

\section{ conclusion}
We have introduced an optical setup for fusing two W states without qubit loss. That is, the present scheme can fuse an $n$-qubit W state and a $m$-qubit W state to get an $(n + m )$-qubit W state. With the assistance of weak cross-Kerr nonlinearities, three fusion steps, two polarization entanglement processes and one spatial entanglement process, are applied, and at last a success probability $\frac{n+m}{2nm}$ can be achieved. Moreover, there is no complete failure output in our QLF fusion scheme, and all the garbage states are recyclable. We have also numerically analyzed the resource cost of the present scheme and the two previous fusion schemes. In addition, the present scheme needs no quantum logic gate and any ancillary photon. Therefore, by virtue of employing the available existing optical elements and mature techniques of measurements, this fusion mechanism is simple and feasible. Furthermore, it may afford the possibility for fusing three or four W states simultaneously.

\vspace{0.5cm}

{\noindent\bf Acknowledgments}\\[0.2cm]

 This work was supported by the National Natural Science Foundation
of China under Grant Nos.11475054, 11371005, Hebei Natural Science Foundation
of China under Grant No.A2016205145, and the Education Department of Hebei Province Natural Science Foundation under Grant No.QN2017089.


\vspace{0.5cm}

\end{document}